\documentclass [a4paper]{IEEEtran}
\usepackage{amsmath}
\usepackage{amsthm}
\usepackage{amssymb, amsfonts}
\usepackage{algorithm}
\usepackage{algorithmic}
\usepackage{graphics}
\usepackage{graphicx}
\usepackage{lipsum}
\usepackage[utf8]{inputenc}
\usepackage{skt}

\renewcommand{\u}[1]{\mathrm{#1}}
\newcommand{\kw}[1]{\mathrm{#1}}

\newcommand{\E}{\mathrm{E}}

\newcommand{\Cov}{\mathrm{Cov}}

\newtheorem{theorem}{Theorem}
\newtheorem{corollary}[theorem]{Corollary}
\newtheorem{lemma}[theorem]{Lemma}
\newtheorem{definition}{Definition}

\title{Linear Extended Whitening Filters}
\author{Aravindh Krishnamoorthy\thanks{\hrule}\thanks{\noindent Aravindh Krishnamoorthy is currently with Ericsson Modem Nuremberg GmbH working in the area of Wireless Communication. E-mail: aravindh.krishnamoorthy@ericsson.com, aravindh.k@ieee.org.}}

\begin{document}
\maketitle

\begin{abstract}
In this paper we present a class of linear whitening filters termed linear extended whitening filters (EWFs) which are whitening filters that have desirable secondary properties and can be used for simplifying algorithms, or achieving desired side-effects on given secondary matrices, random vectors or random processes. Further, we present an application of EWFs for simplification of QR decomposition based ML detection algorithm in Wireless Communication.
\end{abstract}

\section{Introduction}
Whitening filters are widely used across the spectrum of fields for signal whitening or noise pre-whitening where the data is modelled as a random vector or as a wide-sense stationary (WSS) process. 

The whitening process accepts a zero-mean random vector or WSS random process, with a positive-definite Hermitian symmetric covariance matrix as input and produces a zero-mean white vector or process with a covariance matrix $c^2 I$ where $I$ is the identity matrix, for some real valued $c > 0$.

A linear whitening filter is a transformation matrix which transforms the given zero-mean random vector or WSS random process into a white vector. Within the context of this paper, we provide the following definition of terminologies used for whitening filters.

Let $v \in \mathbb{C}^{M}$ be a random vector with zero-mean and positive-definite Hermitian symmetric covariance matrix $\Sigma$, and finite higher-order moments.

\begin{definition}[Linear Whitening Filter]
We define a matrix $F \in \mathbb{C}^{M\text{x}M}$ to be the linear whitening filter (WF) of $v$ if:
\begin{equation}
	F\Sigma F^H = c^2 I_M
\end{equation}
\end{definition}
Where $I_M$ is the MxM identity matrix, and $c \in \mathbb{R}$, $c > 0$ is an arbitrary constant. Further, $v' = Fv$ is called the \emph{whitened vector} and its first two central moments are given by $\E(v') = \u{0}$, and $\Cov(v') = c^2 I_M$.

\begin{definition}[Linear Standard Whitening Filter]
 We define a matrix $F \in \mathbb{C}^{M\text{x}M}$ to be the linear standard whitening filter (SWF) of $v$ if:
\begin{equation}
	F\Sigma F^H = I_M
\end{equation}
\end{definition}
The first two central moments of the whitened vector in this case are given by $\E(v') = \u{0}$, and $\Cov(v') = I_M$.

In the remainder of this paper, without loss of generality, we restrict ourselves to SWFs and transformations of random vectors. However, a parallel of the described properties and methods to WFs and WSS random processes can be easily drawn.

It is well-known that transforming an SWF with an Orthonormal matrix does not change the first two central moments of the whitened vector \cite{paper:eldar}, and therefore an infinite number of SWFs may be developed for a given random vector. However, no statements about comparative performance or equivalence of performance of these SWFs are generally known.

In the first part of this paper, we present a few properties of whitening filters. In specific, we show the relationship between widely used whitening filters and further prove that \emph{all} whitening filters of a random vector are related to each other by an Orthonormal transformation matrix.

If we neglect the effect of whitening on higher order moments, all the SWFs of a random vector are equivalent. Under this assumption, we may choose (of the infinite whitening filters) ones that have secondary properties that are advantageous to the problem on hand. MMSE whitening filter and LS whitening filters (\cite{paper:eldar}, \cite{paper:eldar2}) are examples of whitening filters that have desired secondary properties.

We term an SWF to be a \emph{linear extended whitening filter} (EWF) if it has desirable secondary properties when applied to other given matrices, random vectors or WSS random processes. In the second part of this paper (section \ref{sec:ewf}), we develop a few EWFs.  In the third part of this paper (section \ref{sec:ml}), we present an application of EWFs for ML decoding in Wireless Communication.

\subsection{Computation of SWFs}
Two widely used methods for computing an SWF are 1) based on Cholesky decomposition, and 2) based on Eigenvalue decomposition (see \cite{book:golub}, \cite{book:horn}).
Let $\Sigma = LL^H$ (by Cholesky decomposition), the Cholesky decomposition based SWF is given by:
\begin{equation}
	\label{eqn:cwf}
	F_c = L^{-1}
\end{equation}
Let $\Sigma = Q\Lambda Q^H$ (by Eigenvalue decomposition), the Eigenvalue decomposition based SWF is given by:
\begin{equation}
	F_v = \Lambda^{-1/2} Q^H
\end{equation}

Throughout this paper we use $F_c$ and $F_v$ to refer to Cholesky and Eigenvalue decomposition based SWFs respectively.

\section{Properties of Standard Whitening Filters}
\label{sec:pwf}

In this section we present a few properties of linear standard whitening filters.

\vspace{2mm}
\begin{lemma}
	SWF $F$ is a full-rank matrix.
\end{lemma}
\begin{IEEEproof}
	We have $\kw{rank}(F\Sigma F^H)  = \kw{rank}(I_M) = M$. The equation holds only when $F$ is a full-rank matrix.
\end{IEEEproof}

This lemma ensures that SWFs are invertible. Further, this leads to an interesting result about the mean of input random vector.

\vspace{2mm}
\begin{corollary}
	Let $\mu \ne \u{0}$ be the mean of the random vector $\hat{v} = v + \mu$, there is no SWF $F$ of $\hat{v}$ such that the mean of the transformed vector is $\u{0}$.
\end{corollary}
\begin{IEEEproof}
	The mean of transformed vector is $F\mu$. If $F\mu = \u{0}$ then $\mu$ must lie in the null-space of $F$, which is trivial. Therefore such an SWF does not exist.
\end{IEEEproof}

\vspace{2mm}
Below, we present the well-known property that transforming an SWF by an Orthonormal matrix does not change its whitening properties, i.e. the first two central moments of the whitened vector.

\vspace{2mm}
\begin{lemma}
\label{lm:eq}
	Let $F$ be an SWF of $v$ and $Q$ be an Orthonormal matrix such that $QQ^H = Q^HQ = I_M$, then the matrix $W = Q F$ is also an SWF.
\end{lemma}
\begin{IEEEproof}
	Let $v' = Wv$, we have $\E(v') = 0$ and $\Cov(v') = W\Sigma W^H = Q F \Sigma F^H Q^H = QQ^H = I_M$.
\end{IEEEproof}

\vspace{2mm}
We now show that all SWFs of a random vector are related to each other by an Orthonormal transformation by showing that all the SWFs are related to the Cholesky decomposition based SWF ($F_c$) by an Orthonormal transformation.

\vspace{2mm}
\begin{theorem}
\label{th:rwf}
Every SWF $F$ of $v$ can be expressed as $Q F_c$, where $Q$ is an Orthonormal matrix and $F_c$ is the Cholesky decomposition based SWF.
\end{theorem}
\begin{IEEEproof}
	We have $F^{-1} (F^{-1})^H = \Sigma$. Let $(F^{-1})^H = QR$ (by QR decomposition with positive diagonal elements for $R$), then $R^H Q^H Q R = R^H R = \Sigma$. 
	We have $R^H = L = F_c^{-1}$ (from equation \ref{eqn:cwf}) as the unique Cholesky decomposition \cite{book:golub} of $\Sigma$. 
	Therefore, $F = Q F_c$.
\end{IEEEproof}

\vspace{2mm}
Below we show specific results for relationship between $F_v$ and $F_c$. Similar relationships between $F_v$ and the MMSE whitening filter \cite{paper:eldar} can be shown trivially.

\vspace{2mm}
\begin{corollary}
\label{co:rcewf}
The Eigenvalue decomposition and Cholesky decomposition based SWFs given by $F_v$ and $F_c$  are related as $F_v = Q  F_c$, where $F_v = QR$ (by QR decomposition with positive diagonal elements for $R$).
\end{corollary}
A proof for Corollary \ref{co:rcewf} can be derived along the lines of Theorem \ref{th:rwf}.

\section{Extended Whitening Filters}
\label{sec:ewf}

In this section we develop a few EWFs that have desirable secondary properties when applied to a second random vector, and given second matrix. The EWFs developed for random vectors are equally applicable to WSS random processes.

The general method of finding an EWF is as follows:
\begin{enumerate}
\item Develop an SWF for the given input vector. 
\item  Apply the SWF on a second random vector or matrix.
\item Identify an Orthonormal matrix which produces desired results when applied on the transformed second random vector or matrix.
\item Obtain the EWF by transforming the SWF by the identified Orthonormal matrix.
\end{enumerate}
This method can be easily extended or adapted to develop other EWFs.

\vspace{2mm}
Below, we present an EWF $W$ that is the SWF of first random vector $v$ and de-correlates the second random vector $v_1$ so that $\Cov(Wv_1) = D$, a diagonal matrix.

\vspace{2mm}
\begin{theorem}
	Let $v_1$ be a random vector with mean $\mu$ and positive-definite Hermitian covariance matrix $\Delta$, $F$ be an SWF of $v$, and $Q$ be the Orthonormal matrix such that $F\Delta F^H = Q\Lambda Q^H$ (by Eigenvalue decomposition), then $W = Q^HF$ is the EWF that is the SWF of $v$ and de-correlates $v_1$ such that $\Cov(Wv_1) = \Lambda$.
\end{theorem}
\begin{IEEEproof}
	$W$ is an SWF of $v$ from lemma (\ref{lm:eq}). $\Cov(Wv_1) = W \Cov(v_1) W^H = W\Delta W^H = Q^H F\Delta F^H Q = \Lambda$.
\end{IEEEproof}

\vspace{2mm}
Below, we present two EWFs that triangularize a given second rectangular matrix, or transform a given second square matrix into a positive semi-definite Hermitian matrix. In section (\ref{sec:ml}) we present an algorithmic simplification resulting from the first EWF.

\vspace{2mm}
\begin{theorem}
\label{th:ewfh}
	Let $H \in \mathbb{C}^{M\text{x}N}$ be a rectangular matrix, $F$ be an SWF of $v$, and let $Q$ be the Orthonormal matrix such that $FH = QR$ (by QR decomposition), then the matrix $W = Q^HF$ is the EWF that is the SWF of $v$ and triangularizes the matrix $H$ when applied as $WH$.
\end{theorem}
\begin{IEEEproof}
	$W$ is an SWF of $v$ from lemma (\ref{lm:eq}). We have $WH = Q^HFH = Q^HQR = R$, an upper triangular matrix.
\end{IEEEproof}

\vspace{2mm}
\begin{theorem}
	Let $S \in \mathbb{C}^{M\text{x}M}$ be a square matrix, $F$ be an SWF of $v$, and let $Q$ be the Orthonormal matrix such that $FS = QP$ (by Polar decomposition), then the matrix $W = Q^HF$ is the EWF that is the SWF of $v$ and transforms matrix $S$ into a positive semi-definite Hermitian matrix $P$ when applied as $WS$.
\end{theorem}
\begin{IEEEproof}
	$W$ is an SWF of $v$ from lemma (\ref{lm:eq}). We have $WH = Q^HFH = Q^HQP = P$, a positive semi-definite Hermitian matrix.
\end{IEEEproof}

\section{An Application of EWF}
\label{sec:ml}
In this section we present an application of theorem (\ref{th:ewfh}) for detection in interference for a communication channel using a QR decomposition based ML receiver (along the lines of \cite{paper:kinjo}).

Let $x \in \mathbb{C}^{N_T}$ be the symbols transmitted by $N_T$ antennas and received by $N_R$ antennas and $H \in \mathbb{C}^{N_R\text{x}N_T}$ be the time-invariant channel matrix. The received sequence $y$ may be given as:
\begin{equation}
	y = Hx + v
\end{equation}
Where $v \in \mathbb{C}^{N_R}$ is the interference plus noise (I+N) process with zero-mean and positive-definite Hermitian symmetric covariance matrix $\Sigma$. The I+N process must be pre-whitened before ML detection. Let $F$ be the SWF of $v$, then the model is re-written with white noise as:
\begin{equation}
	\hat{y} = Fy = \hat{H}x + v'
\end{equation}
Where, $\Cov(v') = I_{N_R}$ and $\hat{H} = FH$. The ML estimate of $x$ is given as:
\begin{equation}
	\hat{x} = \arg \min_x ||\hat{y} - \hat{H}x||^2
\end{equation}
QR decomposition based ML detectors use QR decomposition to simplify the ML detection process as follows (e.g. \cite{paper:kinjo}). Let $\hat{H} = QR$ (by QR decomposition), then:
\begin{equation}
	\hat{x} = \arg \min_x ||Q^H\hat{y} - Rx||^2
\end{equation}

\emph{Simplification:} We may use the EWF $W$ developed in theorem (\ref{th:ewfh}) for whitening $v$ instead of an arbitrary SWF $F$. In such a case, $W$ automatically triangularizes $H$ when applied as $WH$ and therefore simplifies the ML detection algorithm by avoiding the filtering of the received signal by $Q^H$. This process is shown below:
\begin{align}
	\hat{y} & = Wy = Rx + v' \\
	\hat{x} & = \arg \min_x ||\hat{y} - Rx||^2
\end{align}
Where $WH = R$, an upper-triangular matrix.

A similar simplification is carried out using EWF $W$ of theorem (\ref{th:ewfh}) for successive interference cancellation in \cite{paper:sim1}.

\newpage

\section{Conclusion}
\label{sec:con}
We presented properties of linear standard whitening filters (SWFs) in section \ref{sec:pwf}, and in specific, relationship between widely used SWFs. 

We presented a method to compute linear extended whitening filters (EWFs) in section \ref{sec:ewf}, and developed a few EWFs. EWFs are SWFs that have desirable secondary properties and can be useful in simplifying algorithms, or achieving desired side-effects on given secondary matrices, random vectors or random processes.

We presented an application of the EWF developed in theorem (\ref{th:ewfh}) for simplification of QR decomposition based ML detection algorithm in Wireless Communication.

\end{document}